# Modeling of hysteresis loop and its applications in ferroelectric materials

Zhi MA*, Yanan MA, Fu ZHENG, Hua GAO, Hongfei LIU, Huanming CHEN*

*School of Physics & Electronic-Electrical Engineering, Ningxia University, Yinchuan 750021, PR China*

*Corresponding author: mazhicn@126.com (Zhi MA), bschm@163.com (Huanming CHEN)

**Abstract.** In order to understand the physical hysteresis loops clearly, we constructed a novel model, which is combined with the electric field, the temperature, and the stress as one synthetically parameter. This model revealed the shape of hysteresis loop was determined by few variables in ferroelectric materials: the saturation of polarization, the coercive field, the electric susceptibility and the equivalent field. Comparison with experimental results revealed the model can retrace polarization versus electric field and temperature. As a applications of this model, the calculate formula of energy storage efficiency, the electrocaloric effect, and the P(E,T) function have also been included in this article.

## 1. Introduction

Due to the intrinsic properties of the ferroelectrics, such as coercive field, spontaneous polarization, and remnant polarizations can be extracted from the polarization-electric field hysteresis loops directly, the polarization-electric field hysteresis phenomena of ferroelectric materials have been investigated extensively since the ferroelectric phenomena being discovered[1-5]. Experimentally, the hysteresis loops can be measured while a periodical electric field or stress being applied onto the ferroelectrics externally[6-7]. Furthermore, many factors, such as temperature, grain boundaries or phase boundaries, doping, and anisotropy etc., which can affect the hysteretic behaviors of the ferroelectrics, have also been studied extensively[8-12]. Theoretically, most of the recent works have been accomplished in model of hysteresis loops by means of the famous Preisach model. However, some effects of microstructure and applied strain or stress taking on the hysteretic behaviors of the ferroelectrics have been predicted for many years similarly, including investigation of the grain size effect on the hysteresis properties of $BaTiO_3$ polycrystals, the effect of dislocation walls on the polarization switching of a ferroelectric single crystal, the effect of strain on the domain switching of ferroelectric polycrystals, and the strain effect in the problem of critical thickness for ferroelectric memory, etc..[13-18]. Many experimental and theoretical results indicated that the hysteretic behaviors of the ferroelectrics are not only dependent on the microstructure of materials (such as grain size, thickness, aging, and grain/phase boundaries) but also dependent on the measurement conditions strongly (such as frequency, temperature and stress etc.)[19-21]. The mechanism of hysteresis loop involved in the ferroelectric system is far from being revealed. Therefore, the reported hysteresis loops measured or predicted under different conditions inevitably presented with diversified patterns. And the diversified patterns obtained experimentally are still beyond on the predictions simulated by using the phenomenological theory of time-dependent Landau Ginzburg Devonshire(TDLGD)[17,18,22].

More recently, Li Jin, etc. summarized the experimental hysteresis loops and categorized them into four groups based on their morphologic features. They also discussed the impact factors on the hysteresis loops[20]. In order to meet with and to understand the diversified patterns of hysteresis loops obtained from experiments theoretically in ferroelectric materials, a new model of hysteresis loops has been derived mathematically in this article. The synthetically parameter (e.g., the combination of electric, temperature, and stress, etc.) employed in this model have been analyzed as one energetic parameter. Meanwhile, a method to calculate the intrinsic parameters of ferroelectric materials has also been given based on our proposed law. Additionally, in order to verify this model, the calculated energy storage efficiency and electrocaloric effect of Pb(Zr$_{0.95}$Ti$_{0.05}$)O$_3$ films have also been included in this article. It will be suggested that the model can give complete intrinsic parameters corresponding to the experiment and which is feasible to describe the diversified hysteresis loops. To the authors' very best knowledge, the unified model proposed for theoretically understanding the diversified patterns of hysteresis loops in ferroelectric materials is first reported.

## 2. Hysteresis loop characters and model in ferroelectric materials

According to the theory of fractal, hysteresis loops or the approximately reverse "*S*" curves obtained from experiments possess the characteristics of fractal. Therefore, we assume that the mathematical model of hysteresis loop can enable us to express its self-similarity. Naturally, based on the notions of self-similar system, let us start from a homogeneous function in a general sense:

$$f(x) = \ln[\cosh(x)] = \ln\left(\frac{e^x + e^{-x}}{2}\right) \tag{1}$$

where $e$ is the base of the natural logarithm. In order to meet with the saturation for both asymptotic values of the externally applied field( $\vec{E}_{app} \to \pm\infty$ ), as well as the behavior of polarization, a differential from the equation (1) is as follow:

$$\frac{df(x)}{dx} = \tanh(x) = \frac{e^x - e^{-x}}{e^x + e^{-x}} = g(x) \tag{2}$$

In order to control and assure the scaling invariant, the function $g(x)$ has to satisfy the following conditions:

$$g(\lambda^\alpha x) = \lambda^\beta g(x) \tag{3}$$

where $\alpha$, $\beta$, and $\lambda$ are the scaling exponents.

We applied the equations mentioned above into the ferroelectric system. The equation (4) describing the polarization-linked free energy density of the ferroelectric system can be constructed, in which the vector $\vec{P}$ is corresponding to the spontaneous polarization, $T$ is temperature, $\sigma$ stands for the stress, and $k$ is polarization corresponding to saturation:

$$F_{\vec{p}}(\vec{P},T,\vec{E},\sigma) = kF(\vec{P},T,\vec{E},\sigma) = k\ln\{\cosh[(aT + b\sigma + m\vec{E} + c)\vec{P} + d] + h\} \tag{4}$$

where, we arbitrarily introduced a linearly action between the spontaneous polarization and the external excitation. For instance, the term $aT$ is the influence of temperature, $b\sigma$ is the linearly contribution of stress, $E$ is the applied electric field. This was inspired by reference [23] which suggested that there is an equivalent effect of the electrical field, the mechanical stress, and the temperature taking on hysteresis loop. The parameters *a*, *b*, *c*, *d*,*m* and *h* are the related parameters during experimental measurement. In order to understand the role of the related parameters during experimental measurement, such as $k$, $d$ and $h$, clearly, we have the derivative of equation (4) with respect to $\bar{P}$ and it can be expressed as equation (5):

$$G(\bar{P},T,\bar{E},\sigma) = k \cdot U \cdot \frac{\sinh(U \cdot \bar{P} + d)}{h + \cosh(U \cdot \bar{P} + d)} \tag{5}$$

Here, we set $U = aT + b\sigma + mE + c$, which contain the contributions to the free energy density that from temperature, the stress, the electric field and etc. It was a Landau-Devoshiro type potential, in which the free energy density is a coupling of the polarizations and the applied external field. It renders that the equation (5) can describe the hysteresis loops of ferroelectric materials. The various patterns of hysteresis loops in ferroelectric materials can be characterized through adjusting the scaling parameters or the synthetically parameter (e.g., electric field, temperature, and stress, etc.) employed in the model. Definitely, we correspond the saturation of polarization, the coercive field, and the slope of the hysteresis loop on the symmetric center to the parameters of $k$, $d$ and $h$ respectively under a constant synthetical conditions. The model of hysteresis loop based on the equation (5) is invariant with respect to scaling and gauge transformation[24], which enables us to express its self-similarity by the homogeneous function in a general sense and reproduce all polarization processes inside a major loop.

## 3. Modeling and simulation of the hysteresis loop

For many years, ferroelectric hysteresis loop has been a subject of extensive research in both experimental and theoretical physics. On one hand, different representatives of the hysteresis loop family exhibit a host of physical phenomena, such as piezoelectricity, ferroelectricity, and electromagnetoelasticity. Ferroelectric materials are used in a broad range of applications, the presence of a hysteresis loop is a basic property characterizing ferroelectric materials. In this sense, hysteresis is unavoidable and must be incorporated in models and subsequent control design to achieve the full capabilities of materials. Simulation of ferroelectric hysteresis loop may aid the design of ferroelectric devices and it may be possible to find an optimum conditions for the ferroelectric system. On the other hand, even though there is a long history of studies of hysteresis loop, building a relatively simple mathematical model of hysteresis loop and determine which factors govern the shape of hysteresis loop is still an attractive challenge for theoretical studies. In order to clarify the hysteresis loop model, we will fit this model to the experimental data with our definitions based on several typical ferroelectric materials. In equation (5), we have defined the *G* function as *P* function in normal hysteresis loops, and the polarization $\bar{P}$ should be replaced as the applied field $E_{app}$ here when the external field was applied since the polarization is determined by the coupling between the temperature field, the electric field and the stress field, now we defined:

$$P = k \cdot U \cdot \frac{\sinh(U \cdot E_{app} + d)}{h + \cosh(U \cdot E_{app} + d)} \qquad (6)$$

here, we set $U = aT + b\sigma + c$, which contains the contributions to the polarizations that from temperature, the stress and etc. Thus, equation (6) has been used to fit the observed hysteresis loops from experimental in BFO($Bi_{0.875}Sm_{0.125}FeO_3$)[25], NBT($Na_{0.5}Bi_{0.5}TiO_3$)[26], BTO($BaTiO_3$)[27] and PZT[28]. If we define the negative coercive field location as the left side of a hysteresis loop, and the positive coercive field location as the right side of a hysteresis loop, the hysteresis loop can be divided into left and right branches. Figure 1 gives the experimental data and the simulated results for the four type ferroelectric materials. The solid lines represent the best simulations for the experimental data. The dotted lines give the measured values from the published papers. The intrinsic parameters in the left loop are consistent with the right ones.

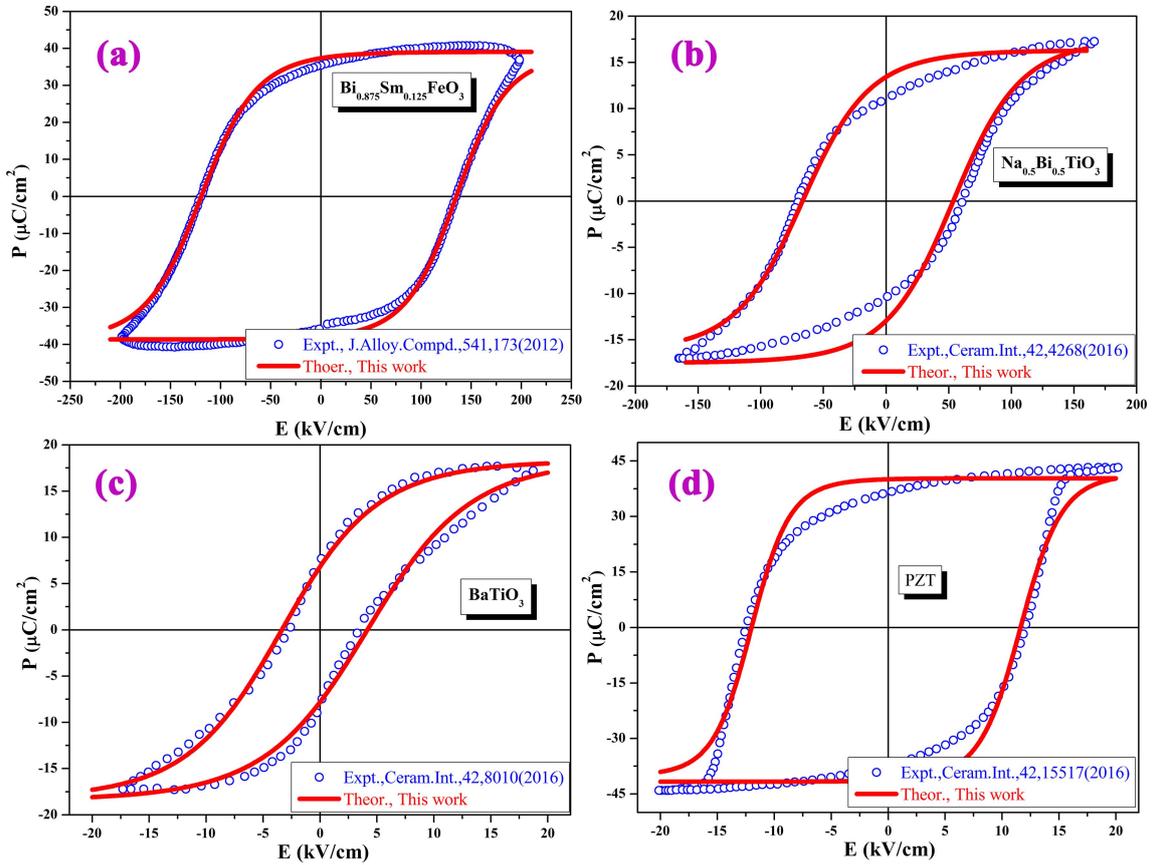

Fig. 1 Fitted result and experimental data for the typical ferroelectric materials
(a)BFO($Bi_{0.875}Sm_{0.125}FeO_3$), (b)NBT($Na_{0.5}Bi_{0.5}TiO_3$), (c)BTO($BaTiO_3$) and (d)PZT

In equation (6), $k$, $U$, $h$ and $d$ are the related parameters during experimental measurement. In previous discussions, we correspond the saturation of polarization, the coercive field, and the slope of the hysteresis loop on the symmetric center to the parameters of $k$, $d$ and $h$ respectively. However, this one to one correspondence is not accurate since the following inequality has been established and proved:

$$k \cdot U \frac{\sinh(U \cdot E_{app} + d)}{h + \cosh(U \cdot E_{app} + d)} \leq |k \cdot U| \qquad (7)$$

Based on the above inequality, $k \cdot U$ has been calculated and compared with the observed $P_s$ values, the experimental data $P_s$ and the theoretical result $k \cdot U$ has been listed in Table 1.

Table 1. The experimental data and theoretical result of intrinsic parameters in several ferroelectric materials

| Parameters | BFO | NBT | BTO | PZT |
|---|---|---|---|---|
| $P_S(\mu C/cm^2)$ | 40.0 | 18.0 | 18.0 | 40.0 |
| $k \cdot U$ | 38.9 | 16.2 | 18.4 | 41.1 |
| $E_C(kV/cm)$ | 130.0 | 60.0 | 3.0 | 12.0 |
| $d/U$ | 127.5 | 62.5 | 3.7 | 11.8 |
| $\chi$ | 0.7 | 0.3 | 2.1 | 11.6 |

From Table 1 we can see that, the calculated values of $k \cdot U$ are in line with the observed parameter $P_s$ in experiment[25-28]. In this sense, $P_s$ can be acquired by directly from equation of $P_S = k \cdot U$. Similarly, $E_C = d/U$, and $\chi = kU^2/(h+1)$, where $\chi$ could be considered as the electric susceptibility when $h$ reflected the slope of the hysteresis loop on the symmetric center. The calculated $d/U$ and $\chi$ values are also listed in Table 1. As is shown in Table 1, all of the calculated values are in consistent with the observed results in experimental measures. Therefore, the equation (6) could be improved as a meaningful hysteresis loop function as following:

$$P = P_S \frac{\sinh(U \cdot E_{app} + U \cdot E_C)}{P_S \cdot U / \chi - 1 + \cosh(U \cdot E_{app} + U \cdot E_C)} \qquad (8)$$

This model implies that the hysteresis loop should be determined by four intrinsic parameters in ferroelectric materials: the saturation of polarization $P_S$, the coercive field $E_C$, the electric susceptibility $\chi$ and the equivalent field $U$. It also shows that, the $P_S$ is proportional to the equivalent field, $E_C$ is inversely proportional to the equivalent field, and the $\chi$ is proportional to the square of the equivalent field. On the other hand, the saturation of polarization $P_S$ and the coercive field $E_C$ will have the following relationship:

$$P_S \cdot E_C = kd \equiv C_1 \qquad (9)$$

The above equation shows that, for a certain ferroelectric materials, it will have an intrinsic constant parameter oneself, this intrinsic constant parameter is independent of any external field. Furthermore, a new physical law can be established between these intrinsic parameters:

$$\frac{\chi E_C}{P_S} = \frac{d}{h+1} \equiv C_2 \qquad (10)$$

This equation shows that the intrinsic parameters of ferroelectric materials are mutually restricted, not independent of each other. According to this model, the intrinsic parameters of ferroelectric materials measured in the experiment have to obey this law.

## 4. The energy storage density and energy storage efficiency

Dielectric materials are used to control and store electric energies and play a key role in modern electronics and electric power systems. Among various dielectric materials, ferroelectrics are presently the material of choice for energy storage applications due to their relatively high energy density, high electric breakdown field, low dielectric loss, fast discharge speed, low cost and high reliability. For the requirements of low cost and high energy storage systems (i.e., the energy density capacitors), the development of high power and energy density materials becomes a major enabling technology. According to the definition of energy storage density in $P-E$ loops, the energy storage efficiency can be calculated by [29]:

$$\eta = \frac{w_1}{w_1 + w_2} \times 100\% \tag{11}$$

where $w_1$ is the recoverable(or discharged) energy density, $w_2$ is the loss energy density. The discharged energy density $w_1$ could be calculated by the discharged loop ($P_L$ represented the left loop)[30]:

$$w_1 = \int_{P_r}^{P_{max}} EdP = \int_0^{E_{max}} (P_{max} - P_L) dE = E_{max} \cdot P_{max} - \int_0^{E_{max}} P_L dE \tag{12}$$

The loss energy density could be calculated by the charged loop ($P_R$ represented the right loop), and then the energy storage efficiency was written as:

$$\eta = \frac{E_{max} \cdot P_{max} - \int_0^{E_{max}} P_L dE}{E_{max} \cdot P_{max} - \int_{E_C}^{E_{max}} P_R dE} \tag{13}$$

Based on the model equation (8), the integral could be calculated as following:

$$\begin{cases} \int_0^{E_{max}} PdE = \frac{P_s}{U} \log \left[ \frac{\cosh(U \cdot E_{max} + U \cdot E_c) + \frac{P_s U}{\chi} - 1}{\cosh(U \cdot E_c) + \frac{P_s U}{\chi} - 1} \right] \\ \int_{E_C}^{E_{max}} PdE = \frac{P_s}{U} \log \left[ \frac{\cosh(U \cdot E_{max} + U \cdot E_c) + \frac{P_s U}{\chi} - 1}{\cosh(2U \cdot E_c) + \frac{P_s U}{\chi} - 1} \right] \end{cases} \tag{14}$$

As illustrated in Fig. 2, the energy storage efficiency could be calculated conveniently by using the expression (14). From the formula (14), we can see that, few variables ($P_S$, U, $E_C$, $\chi$, $E_{max}$) are required to calculate the energy storage efficiency, it is suitable for coupling with field computation and it is compatible with available circuit simulation software.

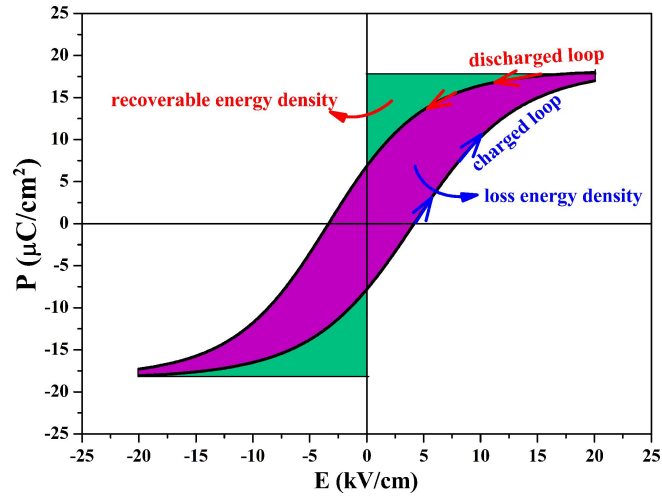

Fig. 2 Diagram of the energy storage during $P-E$ process

## 5. The electrocaloric effect

Ferroelectrics have received much attention in the recent years as potential candidates for solid state colling applications, i.e. using them for refrigeration. The interest in ferroelectrics as potential electrocaloric materials is largely due to the findings of giant electrocaloric effect (ECE) in several ferroelectrics. Refrigeration based on the ECE is more energy efficient and environment friendly and hence offer an alternative to the usual vapor compression refrigeration technology also. In general, the electrocaloric effect is defined as a reversible temperature change in the dielectric materials upon the application of an electric field under adiabatic conditions. The reversible adiabatic change in temperature is given by:

$$\Delta T = -\frac{1}{C \cdot \rho} \int_{E_1}^{E_2} T \left( \frac{\partial P}{\partial T} \right)_E dE \quad (15)$$

where ρ is the materials' density, $C$ is the heat capacity. Usually, when $P(E)$ hysteresis loops were measured under several selected temperatures, the $P(T)$ data at selected applied field could be extracted and then the value of $(\partial P/\partial T)$ can be obtained from $P$ versus $T$ by fitting a fourth order polynomial to the data from the upper branches of hysteresis loops [31]. Based on the Maxwell relation, the isothermal entropy change ΔS due to an applied or removal of an electric field is given by:

$$\Delta S = -\frac{1}{\rho} \int_{E_1}^{E_2} \left( \frac{\partial P}{\partial T} \right)_E dE \quad (16)$$

When the isothermal entropy change ΔS is calculated by equation (16) and recoverable(or discharged) energy density $w_1$ is determined by equation (12), the coefficient of performance (COP) and electrocaloric coefficient γ as a function of temperature could be calculated as following:

$$COP = \frac{|\Delta S \cdot T|}{w_1} \quad (17)$$

$$\gamma = \frac{\Delta T}{\Delta E} \quad (18)$$

From equation (15) to (18), we can see that the estimate of (∂P/∂T) is very important, which implied that the $P(T)$ function is the most important factor for the evaluation of ECE. In the literature, Mischenko ever proposed a fourth polynomial method and used in the year of 2006 [31] and Hamad reported a phenomenological model in 2012 [32]. In this article, we proceed from our proposed model to calculate the electrocaloric effect of Pb(Zr$_{0.95}$Ti$_{0.05}$)O$_3$ films. According to equation (6), the temperature effect should be included under a selected field $E_1$ with a stress free condition:

$$P = K \cdot (AT+B) \cdot \frac{\sinh[(AT+B)\cdot E_1 + D]}{H + \cosh[(AT+B)\cdot E_1 + D]} \quad (19)$$

where, $A$, $B$, $D$, $H$ and $K$ are the four parameters, which is corresponding to the coefficient of the fourth order polynomial model. Equation (15) and (16) are similar, they are both the selected applied field $E$ and temperature $T$ functions. Fig. 3 gives the temperature change at selected values of $\Delta E$ =723 kV/cm. The symbol of star represent the results from Ref. [33], the dashed line is a fourth-polynomial method result and the single line represent simulated data using this model. It is found that, the maximum of temperature change ($\Delta T$) is much lower than that reported in work [33], but close to the one reported in Science where a fourth-polynomial was used [31]. In addition, the corresponding maximum of temperature change ($\Delta T$) in our work are near the experiment result, i.e. the Curie temperature point($T_C$=27℃)[33]. We can see that, in this model, the ECE will not be overestimated and it is more sensitive to the change of the intensity of polarization, which can faithfully reflect the phase transition temperature point.

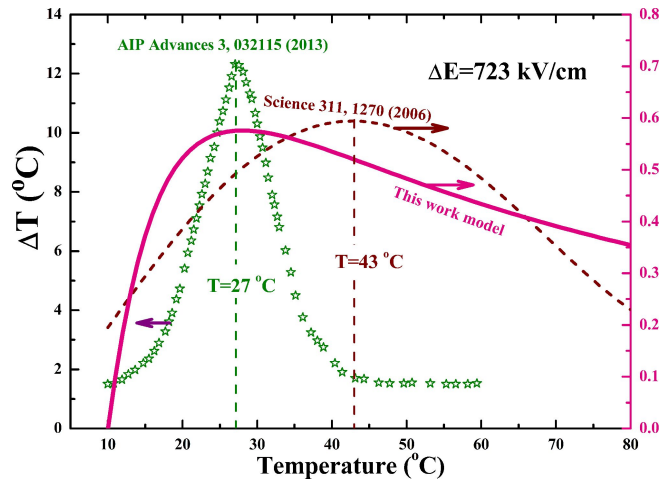

Fig. 3 Electrocaloric temperature changes with $\Delta E$ =723 kV/cm.

## 6.The temperature and electric field effect

It is well known that ferroelectric materials respond linearly at low external fields but evince a nonlinear behavior at high external fields. When the ferroelectric actuators get operated to high temperature environment with much strong electric fields, the temperature and electric field effect should been considered since the performance of piezoelectric devices could been strongly affected by the external fields. Therefore, it is necessary to develop the constitutive model that can predict the nonlinear behaviour of the ferroelectric materials at strong electric field and high temperature.

According to the experimentally measured hysteresis response at different elevated temperatures, the equation (19) has been used to simulate the hysteresis loop since it covered both the electric field effect and the temperature effect[34], which gives the average intrinsic parameters: A=0.00593, B=-3.643, D=3.753, H=0.263, K=0.12. Fig. 4 shows that the polarization is strongly dependent on the temperature and electric field. At low external fields, the polarization is affected stronger by electric field than the temperature. It also illustrates that a narrow strip where polarization is much close to zero will resulted in the system free energy is become much smaller.

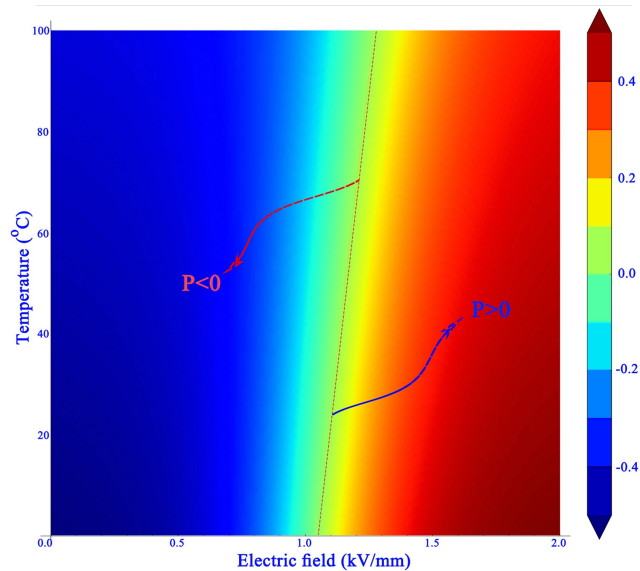

Fig. 4 The density plot of polarization as a function of temperature and electric filed . The dashed line shows where the P=0, the arrows illustrated the change directions of the polarization.

## 7. Summary

In order to meet with and to understand the diversified patterns of hysteresis loops obtained from experiments theoretically in ferroelectric materials, a new model of polarization-electric field hysteresis loops has been derived mathematically. In the proposed model, the external energy, such as the applied electric field, the temperature field, and the stress-strain field, etc., has been analyzed as one energetic parameter synthetically. By using our proposed model, a method to calculate the intrinsic parameters of ferroelectric materials has been given. The results suggested that the model can give a reasonable intrinsic parameter corresponding to the experiment. For validation, four experimental patterns of hysteresis loops reported in literatures have been chosen and verified by our proposed model. The compared results indicated that the model is feasible to describe the diversified hysteresis loops in ferroelectric materials. It can characterize and give complete intrinsic parameter corresponding to the experiment. All these results also revealed that these intrinsic parameters ($P_S$, $E_C$, $U$ and $\chi$) are mutually restricted by the hysteresis loops and obey a simple law, not independent of each other.

As a further verification of this model, the energy storage density, the electrocaloric effect, the applied external field (both the electric filed and temperature) effect have also been disscussed. A parametric study revealed that, the energy storage density is strongly dependent on the electric filed, the electrocaloric effect is strongly dependent on temperature, the polarization is dependent on both the temperature and electric field. Thus the nonlinear constitutive hysteresis model can be efficiently used in designing of ferroelectric devices and predict the non linear behaviour of ferroelectric materials at high electric loading conditions and high temperature.

## Acknowledgements

This work was financially supported by the National Natural Science Foundation of China (NSFC) under Grant number of 11662014, and partially supported by the Major Innovation Projects for Building First-class Universities in China's Western Region (Grant number ZKZD2017006).